\documentclass[a4paper,fleqn]{cas-sc}

\usepackage[numbers,sort&compress]{natbib}

\usepackage{lineno}       
\usepackage{lastpage}     
\AtBeginDocument{}
\usepackage{booktabs}
\usepackage{siunitx}
\usepackage{xcolor}
\newif\ifmarkrevisions \markrevisionsfalse
\ifdefined\RevClean \markrevisionsfalse\fi

\usepackage{multirow}

\graphicspath{{./}}

\begin{document}
\let\WriteBookmarks\relax

\shorttitle{ML for sub-pixel trajectory reconstruction in discretized detectors}
\shortauthors{M.\,M. Romano et al.}

\title[mode=title]{Machine Learning Methods for Sub-Pixel Trajectory Reconstruction in Discretized Position Detectors}

\author[1]{Matthew Mark Romano}[orcid=0000-0002-0612-9003]
\cormark[1]
\ead{mromano3@wisc.edu}
\credit{Conceptualization, Methodology, Formal analysis, Writing -- original draft}

\author[2]{Zhengzhi Liu}
\credit{Software, Validation}

\author[1]{JungHyun Bae}
\credit{Supervision, Writing -- review \& editing, Funding acquisition}

\affiliation[1]{organization={Department of Nuclear Engineering and Engineering Physics, University of Wisconsin--Madison},
            addressline={},
            city={Madison},
            postcode={53706},
            state={WI},
            country={USA}}

\affiliation[2]{organization={Lucid Motors},
            addressline={7373 Gateway Blvd},
            city={Newark},
            postcode={94560},
            state={CA},
            country={USA}}

\cortext[1]{Corresponding author.}

\begin{abstract}
We evaluate machine-learning methods for sub-pixel position and trajectory-angle reconstruction in scintillator detectors. Geant4 simulations use two 50~cm $\times$ 50~cm plastic-scintillator planes separated by 30~cm, each represented by an 8$\times$8 grid of 64 touching computational cells of pitch 6.25~cm. The plates are optically continuous. Photons reaching any outer face are absorbed and scored once in the pre-step cell. We sum the collected photon energies in each cell and normalize the signals to the plane total. Photodetectors and electronics are not simulated. We compare a transformer, a convolutional neural network (CNN), a multilayer perceptron (MLP), and an energy-weighted centroid. We compute directions from the two reconstructed positions. On a common 19{,}053-event test set, transformer and CNN median position errors (per-event means of the two planes' radial errors) are 0.18 and 0.23~cm, or 2.9\% and 3.6\% of the cell width. Their median space-angle errors are 0.35$^\circ$ and 0.52$^\circ$, compared with 1.03~cm and 1.90$^\circ$ for the centroid and 4.21$^\circ$ for an idealized Monte Carlo reference retaining only the traversed cell. The transformer has the lowest median errors, whereas the CNN has the lowest position- and angular-RMSE point estimates. Large-error tails limit the RMSE gains. Learned methods recover sub-pixel information in this idealized response. Establishing gains for muon tomography requires realistic optical coupling, photodetector response, and electronics.
\end{abstract}

\begin{highlights}
\item Machine learning recovers sub-cell positions in optically continuous plates.
\item A hybrid transformer splits reconstruction into cell and in-cell offset estimates.
\item Transformer and CNN cut median position and angle errors 3.7--6.1$\times$ vs. centroid.
\item Simulated transformer median position error is 2.9\% of the 6.25~cm signal-bin pitch.
\end{highlights}

\begin{keywords}
Particle tracking detectors \sep Muon tomography \sep Neural networks \sep Transformer architectures
\end{keywords}

\maketitle

\section{Introduction}
\label{sec:intro}

Muon tomography uses penetrating cosmic-ray muons to image dense materials and large structures~\cite{BONOMI2020,borozdin2003}. Applications include the detection of shielded nuclear materials, geological surveys of volcanic structures, archaeological investigations, and structural monitoring~\cite{BONOMI2020, bae2022}. The field has progressed from early feasibility studies~\cite{alvarez1970} to measurements that resolved meter-scale voids within large structures~\cite{morishima2017}. Further improvements in spatial and angular resolution would extend the range of structures that can be imaged and reduce the exposure required to form a useful tomographic reconstruction.

Muon tomography reconstructs particle trajectories from hit positions measured in multiple detector planes. In a continuous scintillator plate with pixelized collection, a single muon can produce signals in several readout bins. We use this light pattern to infer the crossing point below the bin pitch. As a charged particle traverses a scintillator, it excites the material along its path, and the subsequent de-excitation produces visible light~\cite{Kno2000,dpr1999}. An inclined muon and its secondary particles can deposit energy across several computational cells, and optical photons propagate between the touching PVT placements without internal optical barriers. The collected multi-bin energy pattern carries sub-pitch position information.

A conventional position estimator is the energy-weighted centroid~\cite{RADEKA1980,LANDI2002}:
\begin{equation}
\label{eq:centroid}
\bar{x} = \frac{\sum_{i} E_i x_i}{\sum_{i} E_i} \,, \qquad \bar{y} = \frac{\sum_{i} E_i y_i}{\sum_{i} E_i} \,,
\end{equation}
where $E_i$ is the optical-photon energy collected in computational cell $i$, and $(x_i, y_i)$ are its centre coordinates. The centroid is inexpensive to compute, but its fixed functional form does not capture all spatial correlations in the detector response. It can also produce systematic position errors near bin boundaries, where asymmetric light patterns from inclined tracks and secondary particles can shift the energy-weighted average away from the true crossing point.

Machine-learning methods provide an alternative means of estimating particle position from segmented detector signals~\cite{pnas2018}. Convolutional neural networks (CNNs) learn spatial features from image-like inputs~\cite{krizhevsky2012} and have been applied to sub-pixel reconstruction in pixelated detectors~\cite{VANSCHAYCK2020}. Neural networks have also been used for charged-particle tracking in high-energy physics, including graph neural networks that associate detector hits with candidate tracks~\cite{DeZoort2021}. Transformer architectures were originally developed for sequence modeling~\cite{vaswani2017}. Recent studies have applied transformers to charged-particle tracking in collider detectors, including hit-to-track association and joint hit filtering and track reconstruction~\cite{Caron2025,VanStroud2025}. These methods operate on sparse detector-hit sets and address the combinatorial assignment of hits to tracks rather than sub-pixel position regression from two-dimensional scintillator signal patterns. In scintillator-based muon tomography, sub-pitch resolution has also been demonstrated using triangular strips and charge-centroid reconstruction~\cite{Liang2026}. The MINERvA tracker uses sharing between triangular strips for position interpolation~\cite{Aliaga2014}. Continuous scintillators also use statistics-based maximum-likelihood positioning~\cite{Joung2002} and neural-network estimators~\cite{Bruyndonckx2004}, both requiring a model or calibration of the light-sharing response. Here we compare centroid, MLP, CNN, and classification--offset transformer reconstruction for a coarse 8$\times$8 signal grid on one simulated dataset and held-out test set, then propagate position errors to the two-plane direction. We did not implement a likelihood estimator.

In this work, we investigate whether a transformer can improve position and trajectory reconstruction in a continuous scintillator plate with pixelized readout for muon tomography. The proposed architecture separates the task into coarse cell classification and sub-pixel offset regression. Using cosmic-ray muon events simulated in Geant4 for an 8$\times$8 computational signal grid, we compare the transformer with a CNN, an MLP, and an energy-weighted centroid.

\section{Detector simulation}
\label{sec:simulation}

\begin{figure}[htbp]
\centering
\includegraphics[width=\linewidth]{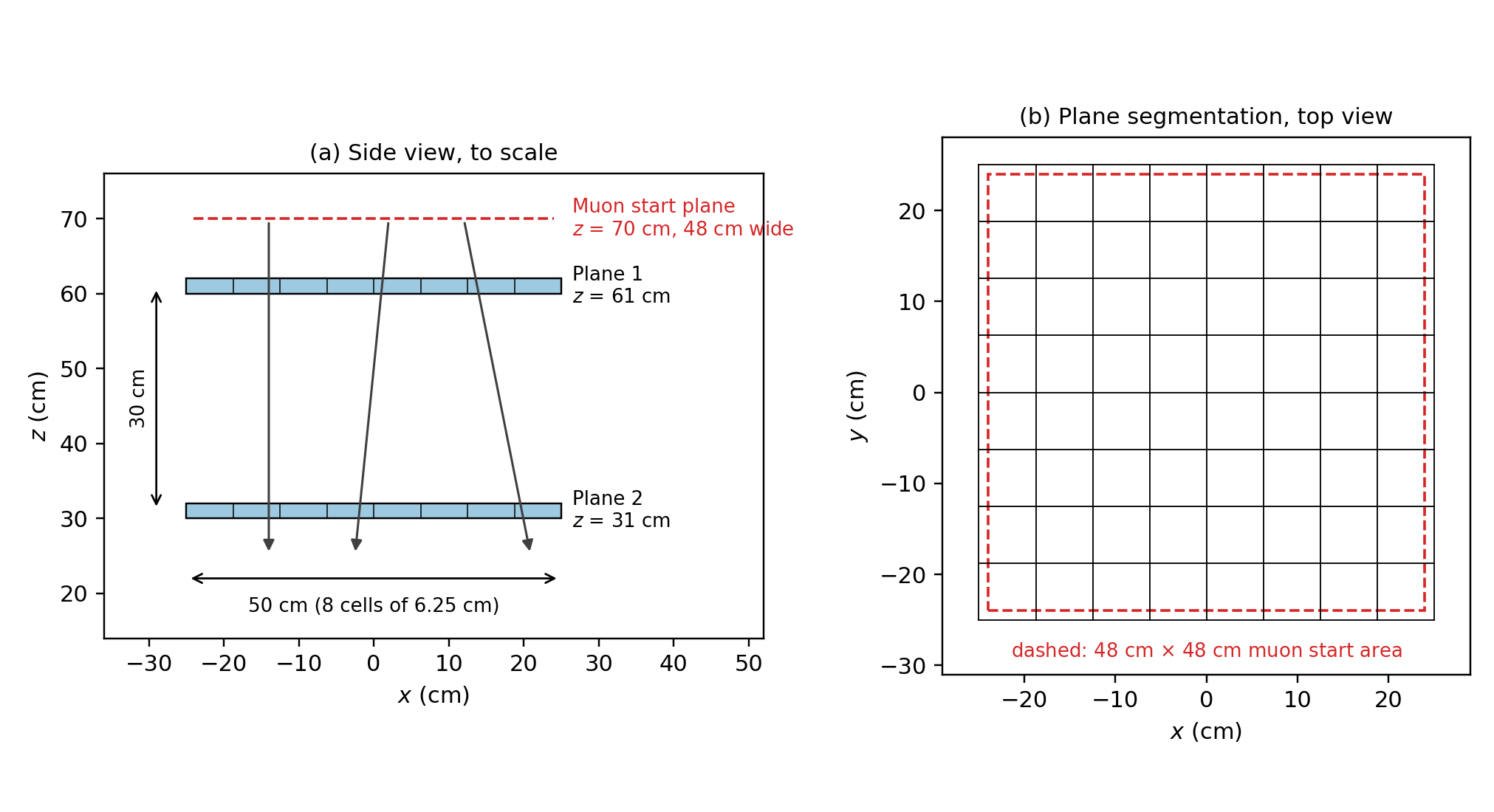}
\caption{To-scale schematic of the simulated detector pair used in this study. (a) Side view: two 2~cm thick PVT planes centred at $z = 61$ and 31~cm, the muon start plane at $z = 70$~cm, and example straight downward tracks at polar angles of 0$^\circ$, 5.7$^\circ$ (the source width $\sigma$), and 11.2$^\circ$ (the 95\% quantile of the source distribution). (b) Top view of the $8 \times 8$ computational signal grid and of the $48 \times 48$~cm$^2$ start area. The grid marks touching PVT placements rather than optical barriers. The simulated geometry also contains a second, identical pair of planes (centred at $z = -31$ and $-61$~cm) that is not used in this work.}
\label{fig:Tomography}
\end{figure}

The Geant4 geometry has four planes in two pairs separated by a 60~cm gap. We analyse only the upper pair, shown to scale in figure~\ref{fig:Tomography}. We reconstruct crossing positions independently in its two planes, separated by 30~cm, and calculate the direction geometrically. Labels were generated only for this pair; the lower pair was not analysed. Readout electronics are not modeled.

We simulated the detector with Geant4 version 11.4.1~\cite{agostinelli2003,eal2016} using the FTFP\_BERT hadronic physics list together with Geant4 optical physics. Each detector plane is a $50 \times 50$~cm$^2$ optically continuous polyvinyltoluene (PVT) scintillator plate represented by an $8 \times 8$ array of 64 touching computational cells. Each computational cell measures $6.25 \times 6.25 \times 2.0$~cm$^3$. Throughout, cell means a computational signal bin rather than an optically isolated scintillator element.

Negative-muon kinetic energies follow a Gaussian centred at 4~GeV with $\sigma = 2$~GeV; values below 1~GeV are set to 1~GeV. The polar angle $\theta$ follows a zero-mean Gaussian with $\sigma = 0.1$~rad, and the azimuth $\phi$ is uniform. Negative draws are equivalent to $|\theta|$ at $\phi + \pi$, giving an effective half-normal distribution with $\sigma = 0.1$~rad (5.7$^\circ$), median 3.9$^\circ$, and 95\% quantile 11.2$^\circ$. The 0.5~rad cap is reached with probability about $6 \times 10^{-7}$ and is negligible. Among the 95{,}262 events crossing both planes, the true polar-angle median, 95\% quantile, and maximum are 3.7$^\circ$, 10.9$^\circ$, and 24.1$^\circ$. The source width, not geometric acceptance, limits this sample: straight tracks can cross both planes at up to 59$^\circ$ edge to edge or 67$^\circ$ corner to corner. This controlled near-vertical source under-represents inclined muons relative to the approximately $\cos^2\theta$ sea-level distribution~\cite{PDG2024}; section~\ref{sec:realism} discusses the consequences. Muons start at $z = 70$~cm, 9~cm above the first plane's centre, uniformly over a centred $48 \times 48$~cm$^2$ area. First-plane crossing coordinates are approximately uniform, with standard deviation 13.5~cm per axis and bounds $\pm$24.6~cm. Truth positions are the starts of the primary muon's first steps inside each plane. The source retains the characteristic cosmic-muon energy scale~\cite{PDG2024}.

The simulation tracks scintillation and optical-photon transport through 64 touching PVT cells per plane, without gaps, wrapping, or internal optical interfaces. Internal same-material crossings neither score nor absorb photons. We score only final exits at the absorbing outer surface of the plate, including its top, bottom, and lateral faces. We add each collected photon's energy in MeV once to its pre-step cell, then normalize the summed cell energies by their total on each plane. This is ideal pixelized collection from a continuous plate, not isolated scintillator cells or bottom-face-only readout. Both planes have a median of 24 cells above 0.1\% of the plane signal. The simulated response has not been compared with measurements, and photodetectors and electronics are not modeled. The effective yield is 1000 photons/MeV, approximately 10\% of the intrinsic PVT yield of $\sim$10{,}000 photons/MeV, representing collection and detection losses rather than a calibrated efficiency. Events require primary-muon intersections with both selected planes; their simulated crossing positions define the true trajectory and angle.

We generated 100{,}000 muon events, of which 95{,}262 intersected both selected detector planes. These events were divided into training, validation, and test sets using a 64\%/16\%/20\% split, yielding 60{,}967, 15{,}242, and 19{,}053 events, respectively. The validation set was used for neural-network model selection and for early stopping during transformer training. All four reconstruction methods were evaluated on the same held-out test set, and the two planes of each event were paired by their Geant4 event identifier. Each event contains the simulated muon position on both planes and its 64-bin collected photon-energy pattern.

The 100{,}000-event simulation took 2497.6~s on one CPU thread. The 10{,}000-event comparison uses the EventID $< 10000$ prefix of this simulation, not an independent simulation, with the same source, seeds, and 64\%/16\%/20\% split procedure (1906 test events). Table~\ref{tab:samplesize} compares the samples. Increasing the sample tenfold reduced learned-method median position and polar-angle errors by 13--45\%. The centroid median position error changed by less than 1\%, while its median polar-angle error increased by 7.2\%. The lower errors show gains beyond 10{,}000 events but do not establish saturation at 100{,}000 events; no intermediate or larger samples were evaluated.

\begin{table}[htbp]
\centering
\ifmarkrevisions\color{blue}\fi
\caption{Median errors for the 10{,}000-event prefix and full 100{,}000-event sample from the same simulation. Position is the per-event mean of the two planes' radial errors; $|\Delta\theta|$ is the absolute polar-angle difference. The samples use the same split procedure but different test sets (1906 and 19{,}053 events).\label{tab:samplesize}}
\smallskip
\begin{tabular}{lcccc}
\toprule
 & \multicolumn{2}{c}{\textbf{Median position error (cm)}} & \multicolumn{2}{c}{\textbf{Median $|\Delta\theta|$ (\si{\degree})}} \\
\textbf{Method} & \textbf{10k run} & \textbf{100k run} & \textbf{10k run} & \textbf{100k run} \\
\midrule
Transformer & 0.209 & 0.181 & 0.260 & 0.210 \\
CNN         & 0.302 & 0.225 & 0.418 & 0.302 \\
MLP         & 0.518 & 0.287 & 0.664 & 0.393 \\
Centroid    & 1.030 & 1.029 & 1.202 & 1.288 \\
\bottomrule
\end{tabular}
\end{table}

\section{Reconstruction methods}
\label{sec:methods}

The MLP, CNN, and transformer use per-plane 8$\times$8 maps of collected optical-photon energy, normalized by the collected energy on that plane. The MLP and CNN were trained with the Adam optimizer~\cite{kingma2017}, a batch size of 64, and an L2 distance loss given by the Euclidean distance between the reconstructed and true positions. The transformer was trained with AdamW and a composite loss containing cell-classification, offset-regression, and absolute-position terms, as described below. It used the same batch size of 64.

For all four methods, the muon direction is obtained geometrically from the reconstructed crossing positions in the two planes and the 30~cm plane separation; no direction information is learned directly. With $\mathbf{d} = (x_2 - x_1,\; y_2 - y_1,\; \Delta z)$ for the reconstructed and true crossing positions, the angular error is quantified primarily by the space angle
\begin{equation}
\alpha = \arctan\!\left(\frac{|\mathbf{d}_\mathrm{rec} \times \mathbf{d}_\mathrm{true}|}{\mathbf{d}_\mathrm{rec} \cdot \mathbf{d}_\mathrm{true}}\right)
\label{eq:spaceangle}
\end{equation}
between the reconstructed and true directions. We also report the polar-angle difference $\Delta\theta = \theta_\mathrm{rec} - \theta_\mathrm{true}$, which is insensitive to first order to position errors perpendicular to the track azimuth and is therefore smaller than $\alpha$. As a reference for the absolute angular scale, we include an idealized hit-cell-only estimate in which each true crossing position is replaced by the centre of the traversed cell. This reference is computed from the Monte Carlo truth; it is not a measured detector baseline and assumes that the traversed cell is always identified correctly. For independent, uniformly distributed quantization errors, this corresponds to a projected-angle standard deviation of $\sqrt{2}\,(w/\sqrt{12})/\Delta z \approx 4.9^\circ$ for cell width $w = 6.25$~cm.

\subsection{Centroid method}

The centroid baseline applies equation~\ref{eq:centroid} independently to each detector plane. This estimator assumes that the recorded signal is distributed symmetrically around the muon crossing point. Near bin boundaries, asymmetric or elongated collected-light patterns from inclined tracks and secondary particles can violate this assumption and shift the reconstructed position. The trajectory angle is calculated from the two reconstructed plane positions and the known plane separation.

\subsection{Multilayer perceptron}

The MLP maps the flattened 64-element signal vector to continuous $(x,y)$ coordinates. Its architecture is
\[
\text{Flatten}
\rightarrow \text{Linear}(64,128)
\rightarrow \text{GELU}
\rightarrow \text{Dropout}(0.1)
\rightarrow \text{Linear}(128,64)
\rightarrow \text{GELU}
\rightarrow \text{Linear}(64,2).
\]
This model provides a learned baseline without imposing an explicit two-dimensional spatial structure on the input.

The MLP was trained for 500 epochs with a learning rate of $1 \times 10^{-3}$ and L2 weight decay of $1 \times 10^{-5}$. The checkpoint with the lowest validation distance error was retained for evaluation. The trajectory angle is calculated geometrically from the positions reconstructed on the two planes.

\subsection{Convolutional neural network}

\begin{figure}[htbp]
\centering
\includegraphics[width=\linewidth]{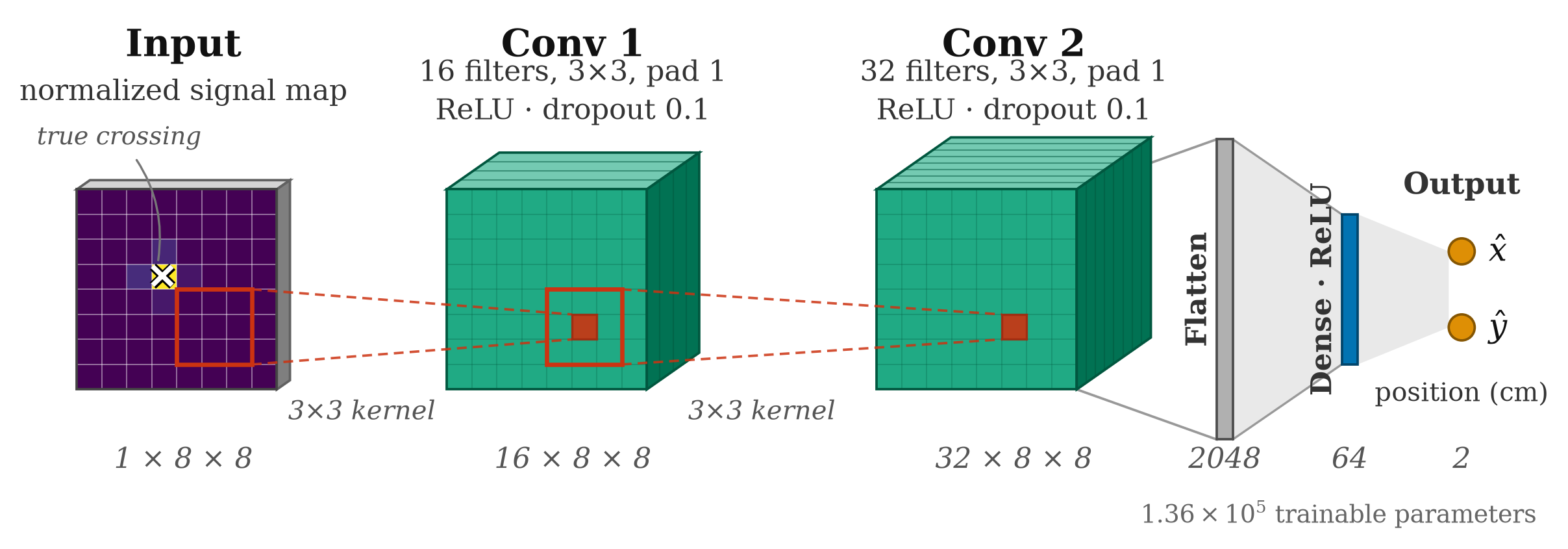}
\caption{CNN architecture used for position reconstruction. The network receives a normalized 8$\times$8 collected photon-energy map (illustrative input; $\times$ marks the true crossing position) and predicts continuous $(x,y)$ coordinates. Feature-map volumes are drawn with depth proportional to the channel count, and the red window and cell illustrate the 3$\times$3 convolution kernel; padding preserves the 8$\times$8 spatial dimensions.}
\label{fig:CNN}
\end{figure}

The CNN retains the two-dimensional arrangement of the scintillator signal map. It contains two convolutional layers followed by fully connected regression layers, as shown in figure~\ref{fig:CNN}.

The first convolutional layer has 16 filters with 3$\times$3 kernels and ReLU activation. The second has 32 filters with the same kernel size and activation. Padding preserves the 8$\times$8 feature-map dimensions so that information from signal bins at the plate boundary is retained. Both convolutional layers use 10\% dropout. The resulting features are flattened and passed through a 64-unit fully connected layer with ReLU activation, followed by a two-unit output layer for the reconstructed $(x,y)$ position.

The CNN was trained for 120 epochs with a learning rate of $1 \times 10^{-3}$ and L2 weight decay of $1 \times 10^{-5}$. The checkpoint with the lowest validation distance error was retained for evaluation.

\subsection{Transformer neural network}

The hybrid transformer separates position reconstruction into coarse cell classification and in-cell offset regression. The classification head identifies the computational cell containing the simulated crossing point, and the offset head estimates the position within that cell.

A convolutional stem converts the 8$\times$8 signal map into 64 tokens, one per computational signal bin, and learned positional embeddings encode their locations. The tokens are processed by a four-layer transformer encoder with eight attention heads and an embedding dimension of 128. Self-attention allows each token to incorporate information from every cell in the signal map, including nonlocal correlations that are not covered by the receptive field of the two-layer CNN. No ablation separates attention from the classification--offset decomposition, so neither component is individually credited for the performance differences.

The network has two prediction heads: a cell-classification head and an in-cell offset-regression head. The in-cell coordinates are normalized to $[-0.5,0.5]$ cell widths. During training, a third auxiliary term stabilizes the absolute coordinate scale; it is evaluated from the true cell center plus the predicted offset rather than from an independently regressed absolute position. The total loss is
\[
\mathcal{L}
=
\mathcal{L}_{\mathrm{cell}}
+
2\mathcal{L}_{\mathrm{offset}}
+
0.5\mathcal{L}_{\mathrm{abs}},
\]
which gives the offset term the largest weight. Classification uses cross-entropy with label smoothing of 0.05; offset and absolute-position terms use L1 loss. The heuristic weights 2 and 0.5 emphasize sub-cell offsets while retaining an auxiliary absolute-position constraint. Neither systematic optimization nor weight-sensitivity testing was performed.

The transformer was trained with AdamW~\cite{loshchilov2019}, a learning rate of $8 \times 10^{-4}$, weight decay of $1 \times 10^{-5}$, and cosine annealing after a five-epoch linear warmup. Gradients were clipped to a maximum norm of 1.0. Training continued for at most 120 epochs, with early stopping after 10 epochs without improvement in the validation distance error.

\begin{figure}[htbp]
\centering
\includegraphics[width=0.92\linewidth]{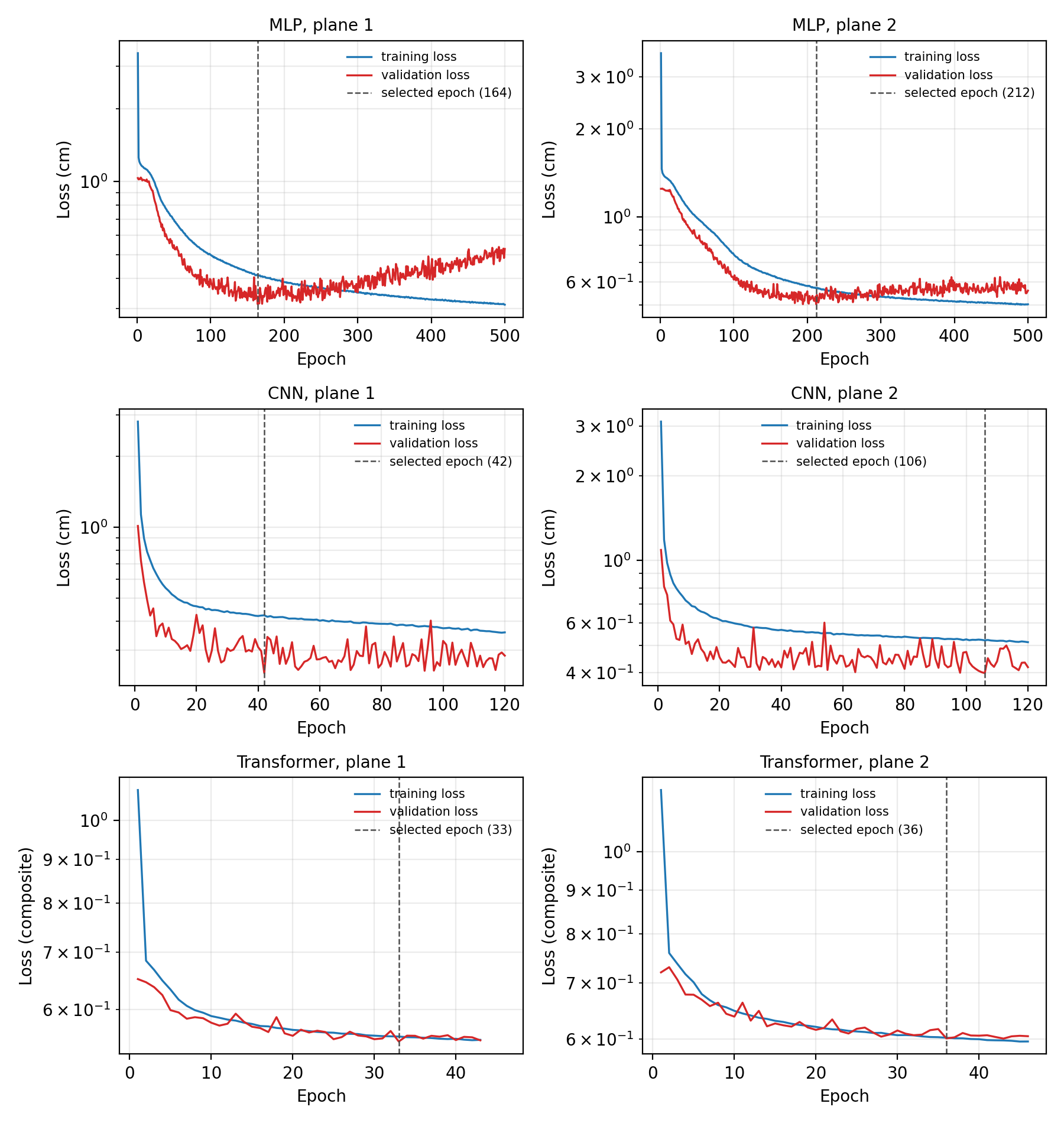}
\caption{Training (blue) and validation (red) loss versus logged epoch for the MLP, CNN, and transformer on each plane. Dashed lines mark the lowest-validation-distance checkpoints used for evaluation. Transformer epoch indices are zero-based. The MLP and CNN loss is the mean Euclidean distance in cm; the transformer loss is the composite loss of section~\ref{sec:methods}. MLP validation loss increases late in training, indicating overfitting beyond the selected checkpoints.}
\label{fig:training}
\end{figure}

Figure~\ref{fig:training} shows the corrected training losses. MLP validation loss rises after its minimum, particularly on plane~1, while training loss continues to decrease. We evaluate the best-validation checkpoints, logged at epochs 164 and 212 for planes~1 and~2. CNN checkpoints are logged at epochs 42 and 106. The transformer selects logged indices 33 and 36 (zero-based) and stops after 44 and 47 epochs. The transformer composite loss includes label-smoothed cross-entropy and is not a distance. Measured training-stage wall times, including startup, evaluation, and output generation, were 765.7 and 762.1~s (MLP), 237.7 and 233.1~s (CNN), and 1163.8 and 1244.7~s (transformer) for planes~1 and~2 on one NVIDIA RTX A1000 GPU. Best-validation weights and launch RNG states were saved. MLP and CNN scripts do not set a training seed, and CUDA operations are not deterministic; no training-seed variance study was performed.

\section{Results and discussion}
\label{sec:results}

We evaluated the four reconstruction methods on the held-out test set of 19{,}053 events, corresponding to 20\% of the 95{,}262 events that intersected both selected detector planes. Tables~\ref{tab:theta_errors}--\ref{tab:perplane} report the angular, position, track-length, confidence-interval, and per-plane results.

\begin{table}[htbp]
\centering\ifmarkrevisions\color{blue}\fi
\caption{Angular reconstruction error statistics for four position reconstruction methods.\label{tab:theta_errors}}
\smallskip
\begin{tabular}{lSSSS}
\toprule
\textbf{Metric} & {\textbf{Transformer}} & {\textbf{CNN}} & {\textbf{MLP}} & {\textbf{Centroid}} \\
\midrule
Mean bias (\si{\degree}) & 0.21 & 0.23 & 0.27 & 0.13 \\
Std deviation (\si{\degree}) & 1.97 & 1.62 & 1.76 & 2.49 \\
RMSE (\si{\degree}) & 1.98 & 1.63 & 1.78 & 2.49 \\
Median $|$error$|$ (\si{\degree}) & 0.210 & 0.302 & 0.393 & 1.288 \\
68th percentile (\si{\degree}) & 0.32 & 0.46 & 0.59 & 2.05 \\
95th percentile (\si{\degree}) & 0.91 & 1.21 & 1.65 & 4.01 \\
99th percentile (\si{\degree}) & 5.48 & 7.12 & 8.73 & 9.54 \\
\bottomrule
\end{tabular}
\end{table}

\begin{table}[htbp]
\centering\ifmarkrevisions\color{blue}\fi
\caption{Position reconstruction error statistics. The per-event position error is the mean of the radial errors on the two planes, $\bar{r} = (r_1 + r_2)/2$; it is not a pooled per-plane statistic, and per-plane radial errors are given in table~\ref{tab:perplane}.\label{tab:position_errors}}
\smallskip
\begin{tabular}{lSSSS}
\toprule
\textbf{Metric} & {\textbf{Transformer}} & {\textbf{CNN}} & {\textbf{MLP}} & {\textbf{Centroid}} \\
\midrule
Mean (\si{\centi\meter}) & 0.27 & 0.32 & 0.41 & 1.17 \\
Std deviation (\si{\centi\meter}) & 0.67 & 0.53 & 0.58 & 0.69 \\
RMSE (\si{\centi\meter}) & 0.72 & 0.62 & 0.71 & 1.36 \\
Median $|$error$|$ (\si{\centi\meter}) & 0.181 & 0.225 & 0.287 & 1.029 \\
68th percentile (\si{\centi\meter}) & 0.24 & 0.28 & 0.36 & 1.20 \\
95th percentile (\si{\centi\meter}) & 0.48 & 0.59 & 0.91 & 2.43 \\
99th percentile (\si{\centi\meter}) & 2.27 & 2.79 & 3.24 & 3.92 \\
\bottomrule
\end{tabular}
\end{table}

\begin{table}[htbp]
\centering\ifmarkrevisions\color{blue}\fi
\caption{3D track length reconstruction errors.\label{tab:track_length_errors}}
\smallskip
\begin{tabular}{lSSSS}
\toprule
\textbf{Metric} & {\textbf{Transformer}} & {\textbf{CNN}} & {\textbf{MLP}} & {\textbf{Centroid}} \\
\midrule
Mean (\si{\centi\meter}) & 0.026 & 0.019 & 0.020 & 0.029 \\
Std deviation (\si{\centi\meter}) & 0.38 & 0.22 & 0.20 & 0.21 \\
RMSE (\si{\centi\meter}) & 0.38 & 0.22 & 0.20 & 0.21 \\
\bottomrule
\end{tabular}
\end{table}

\begin{table}[htbp]
\centering\ifmarkrevisions\color{blue}\fi
\caption{Headline reconstruction metrics with 95\% bootstrap confidence intervals from 2000 resamples of the 19{,}053-event test set. Values are reported as estimate [lower bound, upper bound]. Intervals describe test-sample uncertainty, not training-seed variability; overlap is not a formal test of equivalence.\label{tab:ci}}
\smallskip
\begin{tabular}{lcccc}
\toprule
\textbf{Metric} & \textbf{Transformer} & \textbf{CNN} & \textbf{MLP} & \textbf{Centroid} \\
\midrule
$\Delta\theta$ RMSE (\si{\degree})        & 1.98 [1.78, 2.18] & 1.63 [1.50, 1.78] & 1.78 [1.67, 1.89] & 2.49 [2.42, 2.57] \\
$\Delta\theta$ median $|$err$|$ (\si{\degree}) & 0.210 [0.206, 0.213] & 0.302 [0.297, 0.308] & 0.393 [0.386, 0.399] & 1.288 [1.260, 1.313] \\
Position $\bar{r}$ RMSE (\si{\centi\meter})  & 0.719 [0.653, 0.791] & 0.622 [0.579, 0.668] & 0.706 [0.672, 0.738] & 1.359 [1.339, 1.380] \\
Position $\bar{r}$ median (\si{\centi\meter}) & 0.181 [0.180, 0.183] & 0.225 [0.223, 0.227] & 0.287 [0.285, 0.289] & 1.029 [1.023, 1.035] \\
\bottomrule
\end{tabular}
\end{table}

\begin{table}[htbp]
\centering\ifmarkrevisions\color{blue}\fi
\caption{Per-plane radial position errors. Plane~1 is the first selected plane traversed by the muon, and plane~2 is the second. The neural methods have similar median errors on the two planes but substantially larger plane-2 RMSE, indicating a tail of large plane-2 errors.\label{tab:perplane}}
\smallskip
\begin{tabular}{lSSSS}
\toprule
\textbf{Metric} & {\textbf{Transformer}} & {\textbf{CNN}} & {\textbf{MLP}} & {\textbf{Centroid}} \\
\midrule
Plane 1 RMSE (\si{\centi\meter})           & 0.24 & 0.29 & 0.37 & 1.16 \\
Plane 1 median $|$error$|$ (\si{\centi\meter}) & 0.166 & 0.212 & 0.275 & 1.016 \\
Plane 2 RMSE (\si{\centi\meter})           & 1.36 & 1.12 & 1.23 & 1.74 \\
Plane 2 median $|$error$|$ (\si{\centi\meter}) & 0.181 & 0.218 & 0.277 & 1.051 \\
\bottomrule
\end{tabular}
\end{table}

\subsection{Angular reconstruction performance}

Table~\ref{tab:spaceangle} lists the space-angle errors. Median errors are 0.35$^\circ$ for the transformer, 0.52$^\circ$ for the CNN, 0.68$^\circ$ for the MLP, and 1.90$^\circ$ for the centroid, corresponding to reductions of 5.4$\times$, 3.7$\times$, and 2.8$\times$ relative to the centroid. All four methods improve on the idealized hit-cell-only reference, which has a median of 4.21$^\circ$. The learned methods' space-angle RMS values, 2.05--2.34$^\circ$, have overlapping bootstrap intervals and are dominated by the large-error tail discussed in section~\ref{sec:tail}. For the learned methods, the ratio of median space-angle to median polar-angle error is approximately 1.7, close to the ratio of 1.75 expected between the medians of an isotropic two-dimensional Gaussian error and its one-dimensional projection.

\begin{table}[htbp]
\centering
\ifmarkrevisions\color{blue}\fi
\caption{Space-angle errors $\alpha$ (equation~\ref{eq:spaceangle}) for the 19{,}053-event test set, with 95\% bootstrap confidence intervals from 2000 resamples. The hit-cell-only row replaces each true crossing position with the centre of the traversed cell and serves as an idealized reference for a readout that identifies only the traversed cell.\label{tab:spaceangle}}
\smallskip
\setlength{\tabcolsep}{4pt}
\begin{tabular}{lcccc}
\toprule
\textbf{Method} & \textbf{Median (\si{\degree})} & \textbf{68\% quantile (\si{\degree})} & \textbf{RMS (\si{\degree})} & \textbf{95\% quantile (\si{\degree})} \\
\midrule
Transformer   & 0.354 [0.350, 0.359] & 0.492 [0.486, 0.499] & 2.34 [2.13, 2.55] & 1.29 \\
CNN           & 0.519 [0.514, 0.525] & 0.693 [0.685, 0.699] & 2.05 [1.91, 2.18] & 1.69 \\
MLP           & 0.677 [0.668, 0.684] & 0.902 [0.892, 0.911] & 2.28 [2.17, 2.40] & 2.62 \\
Centroid      & 1.90 [1.86, 1.93] & 2.70 [2.66, 2.73] & 3.11 [3.03, 3.19] & 4.66 \\
Hit-cell only & 4.21 [4.13, 4.29] & 6.43 [6.36, 6.52] & 5.88 [5.83, 5.93] & 10.8 \\
\bottomrule
\end{tabular}
\end{table}

Polar-angle RMSE values are 1.98$^\circ$ for the transformer, 1.63$^\circ$ for the CNN, 1.78$^\circ$ for the MLP, and 2.49$^\circ$ for the centroid. At full precision, the transformer and CNN bootstrap intervals narrowly do not overlap, although their rounded bounds coincide in table~\ref{tab:ci}; the MLP interval overlaps both. All three learned-method intervals are separated from the centroid interval. These intervals describe test-sample uncertainty for the fitted models, not variation across training seeds.

The methods separate more clearly in median absolute polar-angle error. The transformer reaches 0.21$^\circ$, followed by the CNN at 0.30$^\circ$, the MLP at 0.39$^\circ$, and the centroid at 1.29$^\circ$. These are reductions of 6.1$\times$, 4.3$\times$, and 3.3$\times$ relative to the centroid. Heavy-tailed error distributions explain the different RMSE and median rankings: most events are reconstructed accurately, but a few large errors increase the RMSE.

Mean polar-angle biases range from 0.13$^\circ$ to 0.27$^\circ$. The larger differences among standard deviations and upper percentiles therefore arise primarily from the spread of the error distributions rather than a uniform angular offset.

\subsection{Position reconstruction performance}

The CNN has the lowest position-RMSE point estimate at 0.62~cm, followed by the MLP at 0.71~cm and the transformer at 0.72~cm. The transformer interval overlaps both other learned-method intervals, but the CNN and MLP intervals do not overlap (table~\ref{tab:ci}). All three are separated from the centroid result of 1.36~cm. The comparison applies to these fitted models and this test sample; it does not establish a seed-independent ranking.

Median position errors separate the methods more clearly. The transformer reaches 0.18~cm, followed by the CNN at 0.23~cm, the MLP at 0.29~cm, and the centroid at 1.03~cm. These are reductions of 5.7$\times$, 4.6$\times$, and 3.6$\times$ relative to the centroid. The transformer median is approximately 2.9\% of the 6.25~cm cell width. All three learned models recover sub-cell position information, with the transformer giving the lowest typical-event error.

\subsection{Track length errors}

Track-length reconstruction is a weak discriminator among the methods because the fixed 30~cm plane separation is much larger than the transverse displacement between reconstructed positions. RMSE values are 0.22~cm for the CNN, 0.20~cm for the MLP, 0.21~cm for the centroid, and 0.38~cm for the transformer (table~\ref{tab:track_length_errors}).

All four mean track-length errors are below 0.03~cm in magnitude. The transformer's larger RMSE is associated with its tail of large plane-2 position errors. The centroid's cell-scale position errors have less effect on track length because the longitudinal plane separation dominates the calculation.

\subsection{Error analysis}
\label{sec:tail}

\subsubsection{Error distribution characteristics}

\begin{figure}[htbp]
\centering
\includegraphics[width=\linewidth]{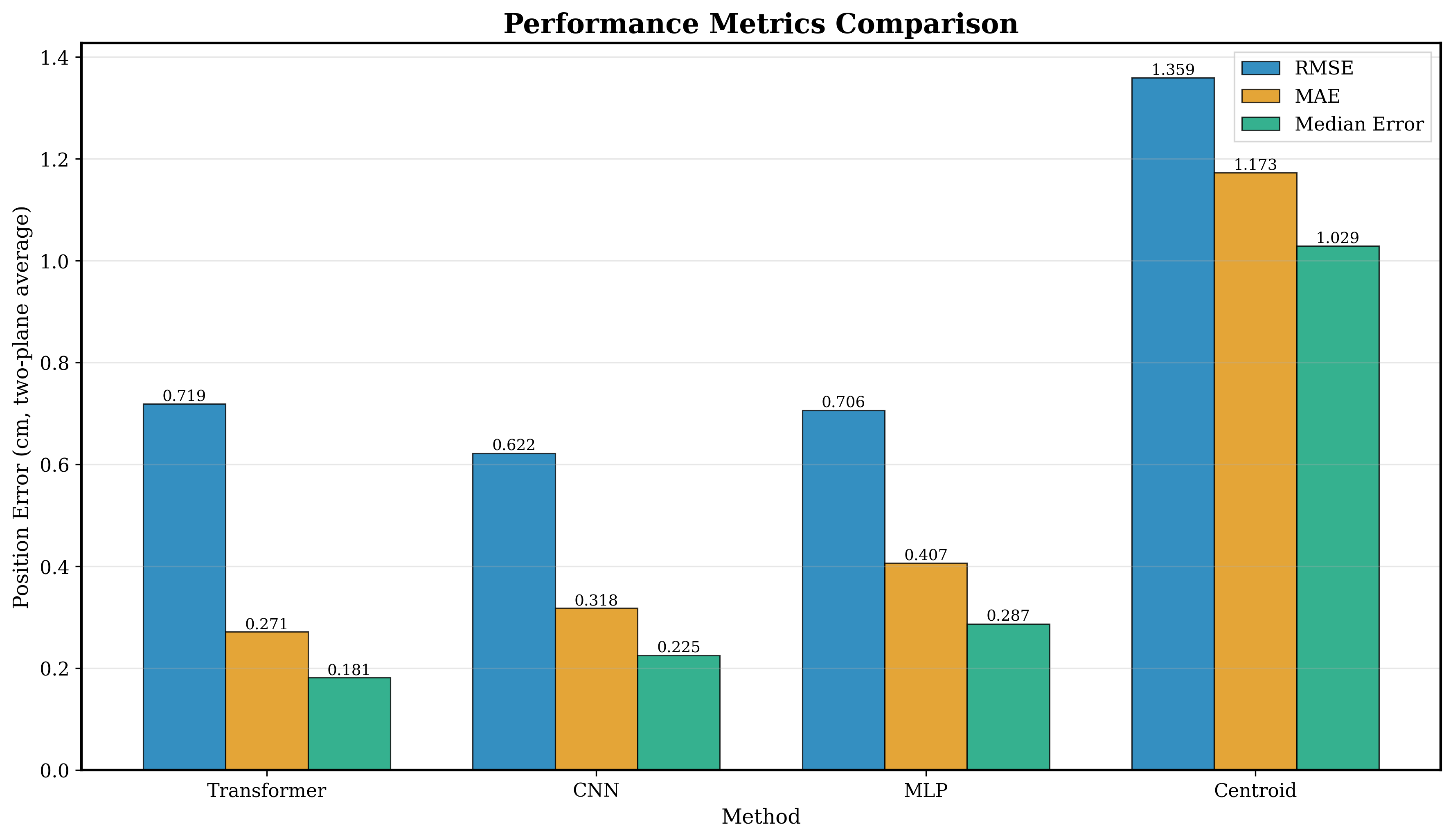}
\caption{Comparison of position-error metrics for the four reconstruction methods. The CNN has the lowest RMSE, while the transformer has the lowest mean and median errors.\label{fig:Performance_Summary}}
\end{figure}

Figure~\ref{fig:Performance_Summary} summarizes the differences between the RMSE and typical-event metrics. The CNN has the lowest position RMSE, while the transformer has the lowest mean and median position errors. The MLP has an RMSE similar to those of the CNN and transformer but a larger median error. All three learned methods outperform the centroid in both RMSE and median position error.

The percentiles expose the large-error tails. For the transformer, 68\% of absolute polar-angle errors are below 0.32$^\circ$ and 95\% are below 0.91$^\circ$, compared with 2.05$^\circ$ and 4.01$^\circ$ for the centroid. At the 99th percentile, the transformer error reaches 5.48$^\circ$, indicating that a small subset of events remains difficult for every reconstruction method.

\subsubsection{Spatial error patterns}

\begin{figure}[htbp]
\centering
\includegraphics[width=\linewidth]{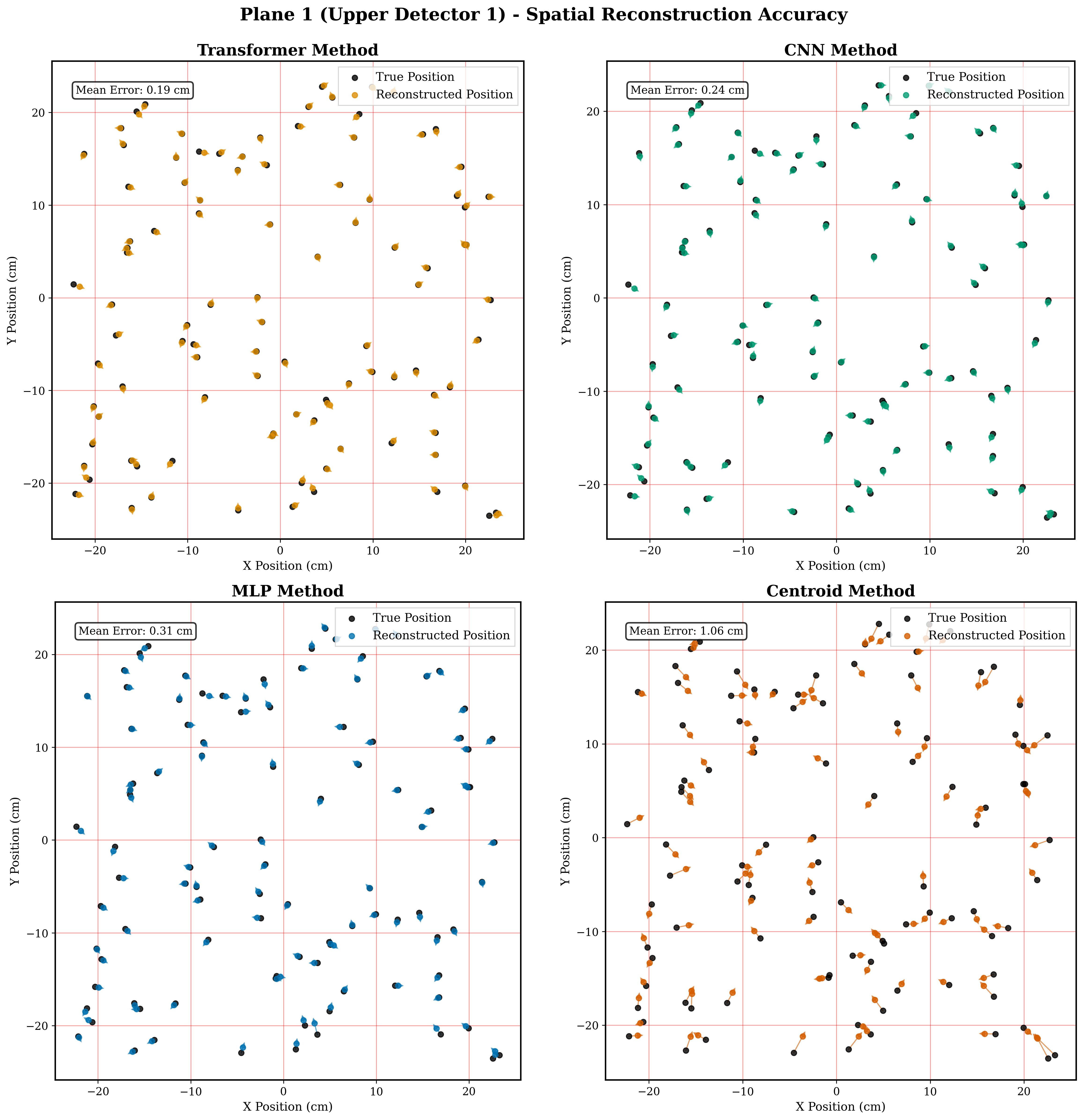}
\caption{Spatial reconstruction errors across detector plane~1. Markers show true muon positions, and arrows connect them to the reconstructed positions for each method.\label{fig:Plane1}}
\end{figure}

\begin{figure}[htbp]
\centering
\includegraphics[width=\linewidth]{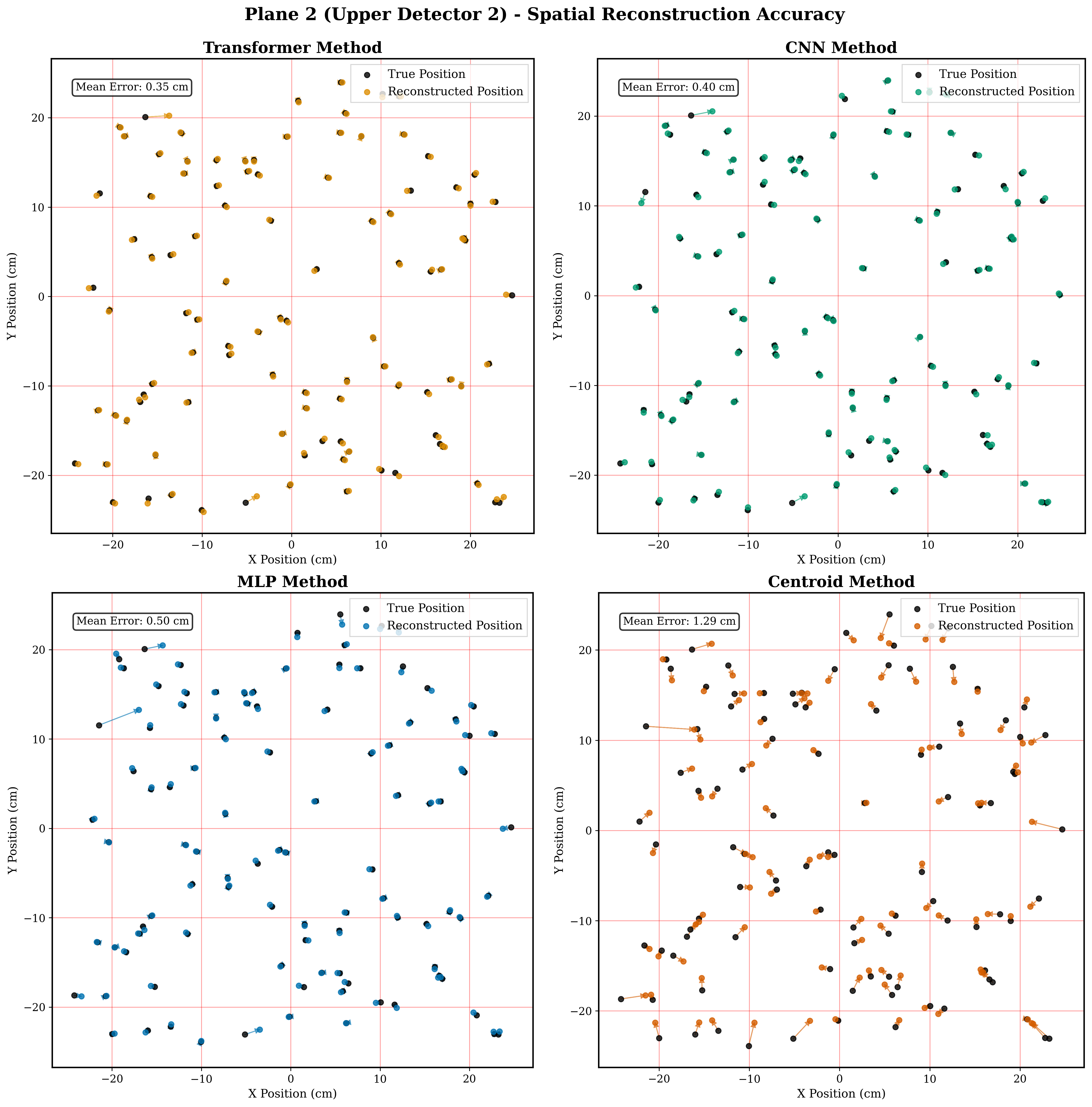}
\caption{Spatial reconstruction errors across detector plane~2. Markers show true muon positions, and arrows connect them to the reconstructed positions for each method.\label{fig:Plane2}}
\end{figure}

Figures~\ref{fig:Plane1} and~\ref{fig:Plane2} show the spatial distribution of the reconstruction errors. On plane~1, the transformer and CNN generally produce shorter error vectors than the centroid method. The centroid errors retain a visible correlation with the computational signal grid, particularly near bin boundaries, where asymmetric collected-light patterns shift the energy-weighted position.

Plane~2 has similar median errors but substantially larger RMSE for the learned methods. This difference indicates that the degradation is concentrated in a small number of events rather than distributed uniformly across the sample.

\begin{figure}[htbp]
\centering
\includegraphics[width=\linewidth]{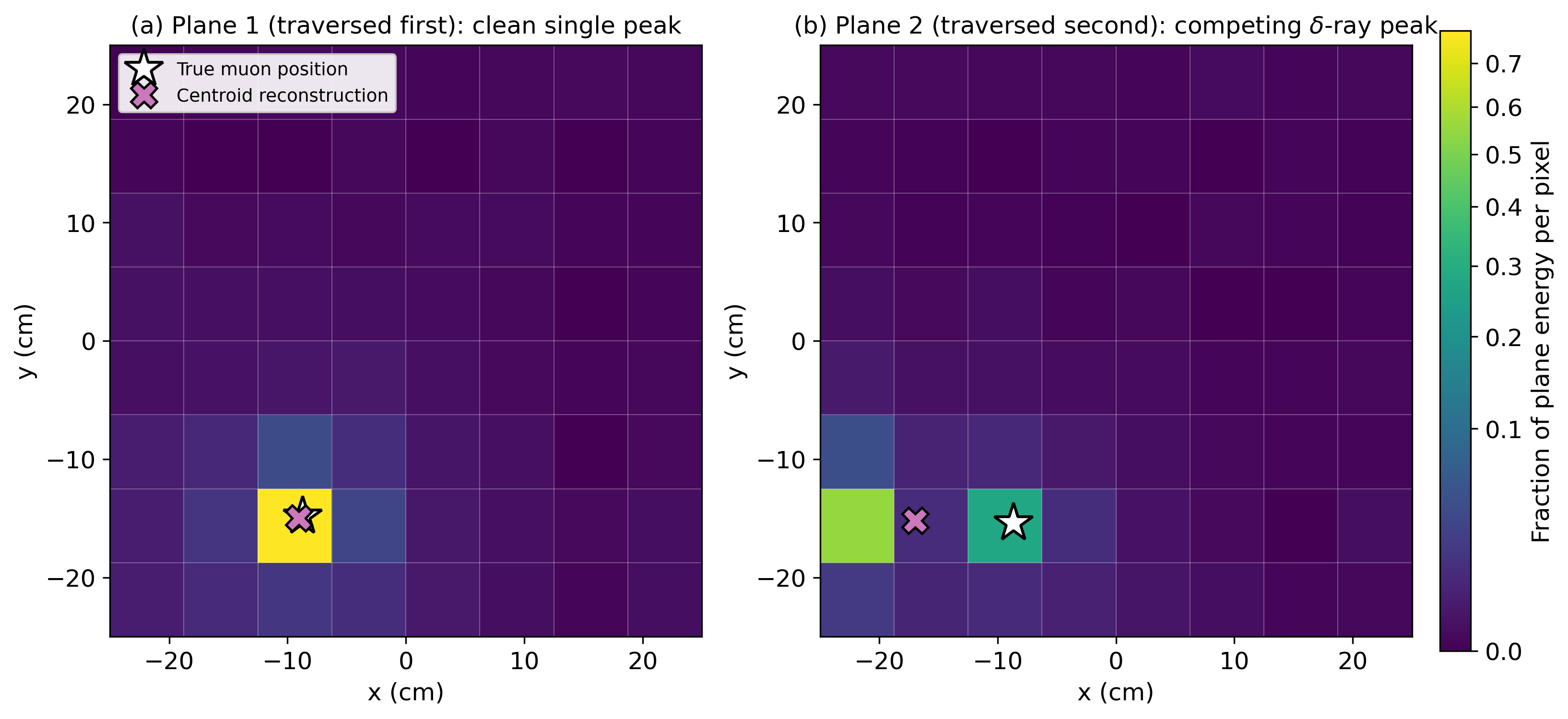}
\caption{Normalized collected photon-energy maps for simulated event 7131 on both planes. The white star marks the true muon position, and the cross marks the centroid reconstruction. The square-root colour scale shows each cell's fraction of the plane's collected photon energy. On plane~2, a competing peak consistent with secondary-particle activity contains 55.3\% of the signal, while the muon cell contains 27.7\%; the peak-cell centre is 13.22~cm from the true crossing point.\label{fig:EventDisplay}}
\end{figure}

The learned methods' plane-2 RMSE values are 1.12--1.36~cm, approximately three to six times their plane-1 values of 0.24--0.37~cm. Their plane-2 medians remain between 0.18 and 0.28~cm, close to the plane-1 medians. The larger RMSE reflects a tail of large errors rather than a uniform loss of precision.

In 0.89\% of plane-2 events, the brightest cell centre is displaced by more than one cell width from the true muon position. In this subset, the muon cell contains a median of approximately 28\% of the collected signal, while a spatially separated competing peak, consistent with secondary-particle activity, appears in a more distant cell. Figure~\ref{fig:EventDisplay} shows one such event. The corresponding condition occurs in 0.001\% of plane-1 events. We compute these fractions over all 95{,}262 labeled events common to both planes, not the held-out test set.

When the strongest signal is spatially separated from the true crossing point, an estimator using only the 64-cell map faces an ambiguous input. This ambiguity contributes to the large-error tail shared by all four methods. Tail-aware training or additional detector information may reduce these errors, but neither was evaluated here.

The error tail explains the differing median and RMSE rankings. Plane~2 errors above half a cell width (3.125~cm) occur in 1.3\% of transformer, 2.0\% of CNN, and 2.6\% of MLP test events, but contribute 95\%, 88\%, and 85\% of their respective plane-2 mean squared errors. Tail RMS errors are 11.8~cm, 7.4~cm, and 7.0~cm; excluding these events gives RMSEs of 0.32~cm, 0.39~cm, and 0.48~cm, respectively. On plane~1, such events constitute less than 0.03\%, and the transformer has the lowest RMSE (table~\ref{tab:perplane}). Fewer than 2.7\% of plane-2 events determine the learned-method RMSE ranking. One possible mechanism, not isolated by ablation, is the transformer's cell-classification--offset decomposition: selecting a competing signal's cell produces a cell-scale position error. Direct CNN regression can instead return intermediate positions, producing more frequent but smaller tail errors. Predictions lie outside the true cell in 3.5\% of transformer versus 4.9\% of CNN plane-2 events. The transformer performs better for typical events but has larger rare failures.

\subsubsection{Angular and position error distributions}

\begin{figure}[htbp]
\centering
\includegraphics[width=\linewidth]{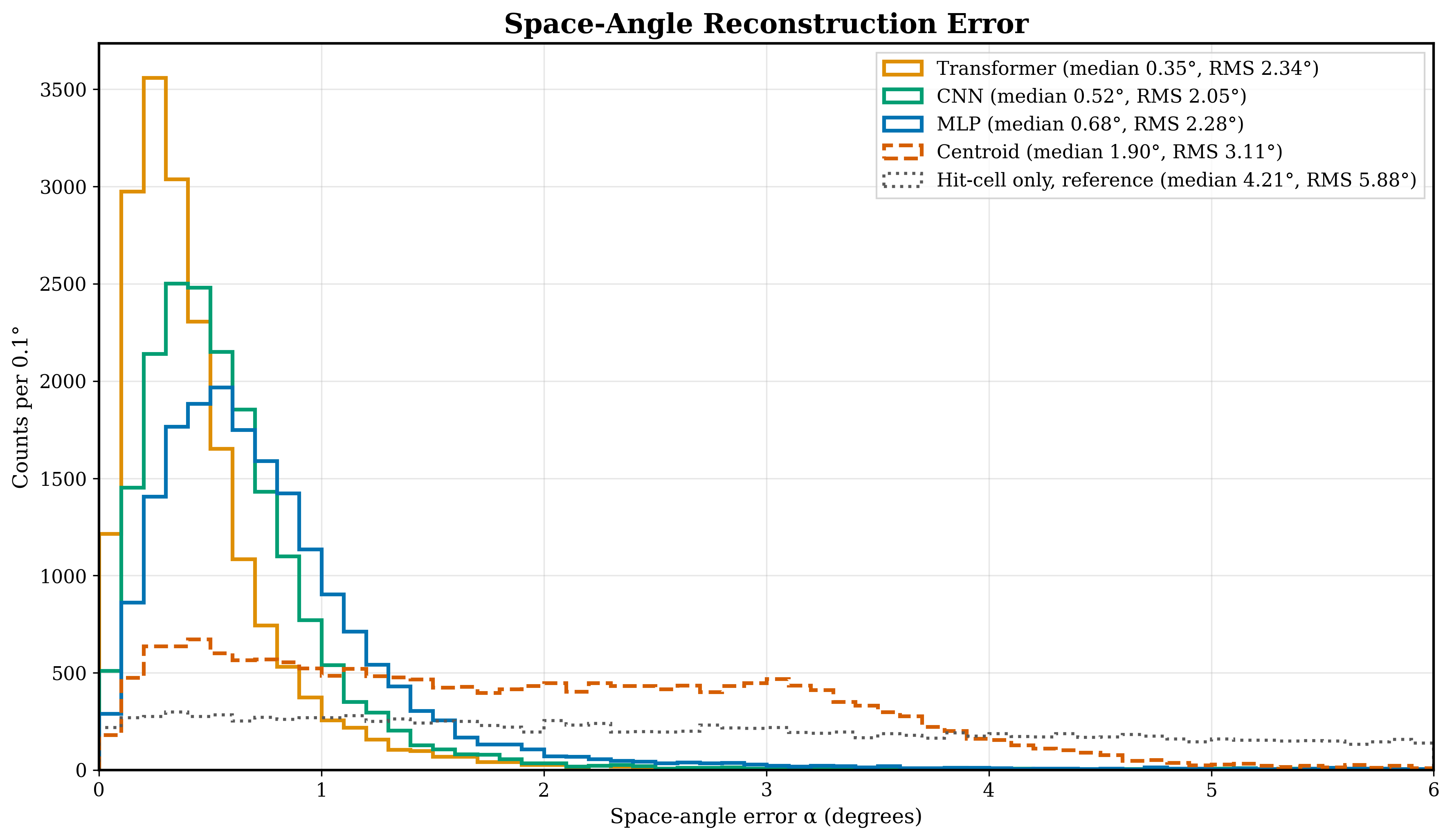}
\caption{Space-angle error ($\alpha$, equation~\ref{eq:spaceangle}) distributions for the four methods and idealized hit-cell-only reference, in 0.1$^\circ$ bins. For the transformer, 95\% of events have errors below 1.29$^\circ$. Errors with $\alpha > 6^\circ$ (1.2\% for the transformer, 2.0\% for the CNN, 2.6\% for the MLP, 3.1\% for the centroid, and 35\% for the hit-cell-only reference) lie outside the plotted range.\label{fig:Angle_Histogram}}
\end{figure}

\begin{figure}[htbp]
\centering
\includegraphics[width=\linewidth]{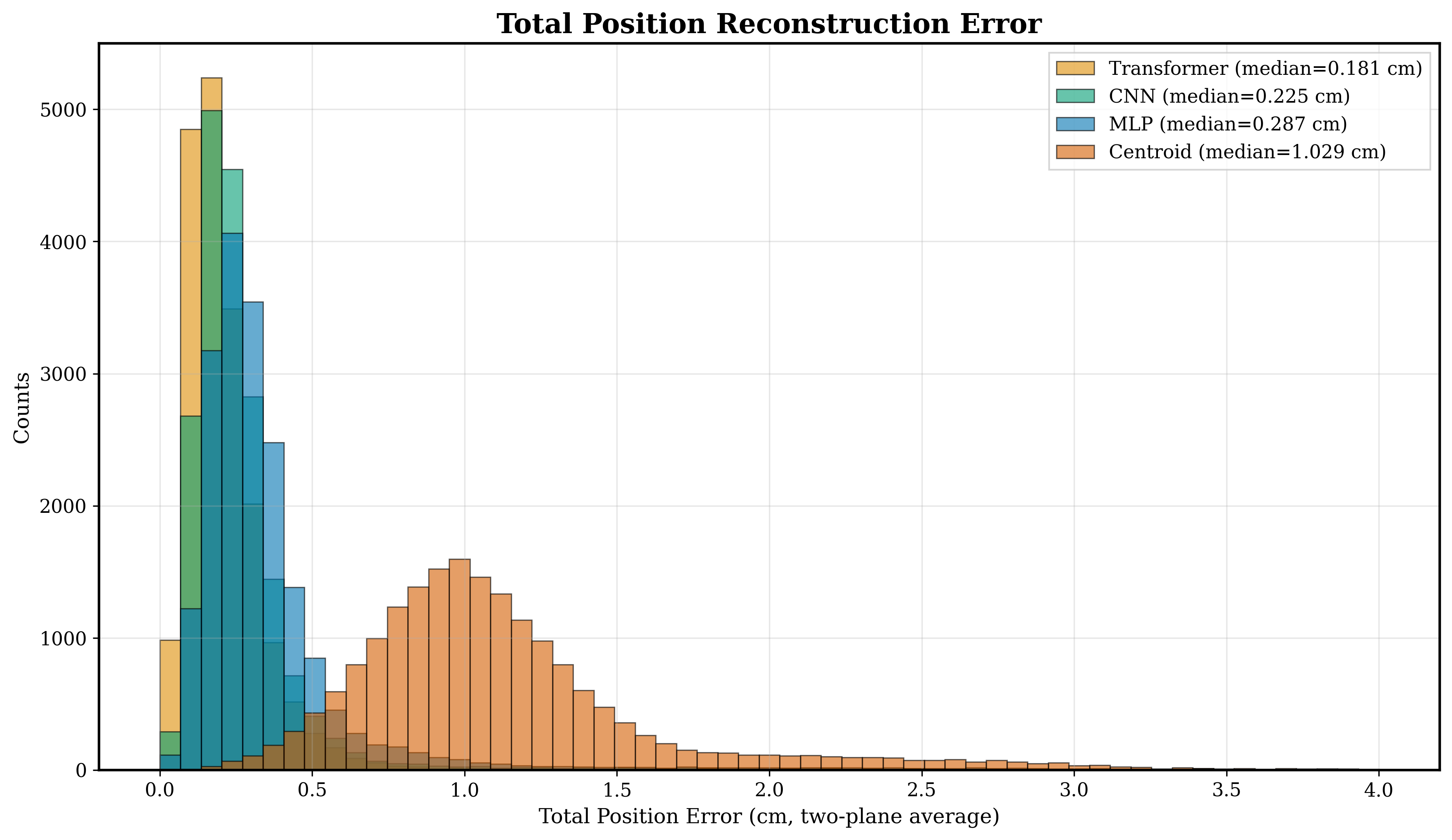}
\caption{Total position-error distributions, using the per-event two-plane mean radial error $\bar{r}$. Median errors are 0.18~cm for the transformer, 0.23~cm for the CNN, 0.29~cm for the MLP, and 1.03~cm for the centroid.\label{fig:Distance_Histogram}}
\end{figure}

The space-angle error histogram (figure~\ref{fig:Angle_Histogram}) shows a narrow central distribution for the transformer, with 95\% of events below approximately 1.29$^\circ$. The broader centroid distribution reflects the propagation of position errors from both detector planes into the reconstructed angle.

The total position-error histogram (figure~\ref{fig:Distance_Histogram}) shows median errors of 0.18~cm for the transformer, 0.23~cm for the CNN, 0.29~cm for the MLP, and 1.03~cm for the centroid. The learned methods recover sub-cell position information for typical events. Image-level improvements were not evaluated in this study, and the heavy error tails remain concentrated on the second selected detector plane.

\subsection{Computational cost}
\label{sec:timing}

Table~\ref{tab:timing} reports inference cost. Best-validation weights from the corrected training runs are saved. The retained timing benchmark was collected separately with the same architectures and untrained weights; it was not rerun with these checkpoints and does not measure accuracy. Inputs are 8$\times$8 production maps. Inference-mode timing uses 20 warm-up calls followed by 300 calls at batch size 1 or 50 at batch size 1024, with GPU synchronization. Transformer timing includes cell selection and offsets. All networks exceed $4 \times 10^3$ events/s on either device at batch size 1024; the transformer at batch size 1 reaches about 500 events/s on the CPU. This exceeds the cosmic-muon rate through 0.25~m$^2$, about 40 events/s at a horizontal flux of $\sim$1~cm$^{-2}$\,min$^{-1}$~\cite{PDG2024}, by at least an order of magnitude. At batch size 1024 on the CPU, the centroid is about 5 times faster than the MLP and 4000 times faster than the transformer.

\begin{table}[htbp]
\centering
\ifmarkrevisions\color{blue}\fi
\caption{Inference time per event (median of the timed calls) for an Intel x86-64 CPU (PyTorch 2.12.1, 20 threads) and an NVIDIA RTX A1000 GPU. The centroid is a vectorized NumPy implementation on the CPU. Network timings use production architectures with untrained weights (see text).\label{tab:timing}}
\smallskip
\begin{tabular}{lrrrrr}
\toprule
 & & \multicolumn{2}{c}{\textbf{CPU (\si{\micro\second}/event)}} & \multicolumn{2}{c}{\textbf{GPU (\si{\micro\second}/event)}} \\
\textbf{Method} & \textbf{Parameters} & \textbf{batch 1} & \textbf{batch 1024} & \textbf{batch 1} & \textbf{batch 1024} \\
\midrule
Centroid    & --      & 7.8  & 0.057 & --   & --   \\
MLP         & 16{,}706  & 116  & 0.31  & 62   & 0.13 \\
CNN         & 136{,}066 & 128  & 1.6   & 124  & 0.72 \\
Transformer & 581{,}637 & 1943 & 226   & 1407 & 92   \\
\bottomrule
\end{tabular}
\end{table}

\subsection{Detector realism and source dependence}
\label{sec:realism}

All methods use the multi-cell patterns produced by optical transport in the continuous plate and idealized pixelized collection (section~\ref{sec:simulation}). Geant4 scintillation emission and optical transport remain stochastic. We do not add a photosensor detection or photoelectron-statistics model, calibrated photon-detection efficiency, gain variation, electronic noise, thresholds, or readout crosstalk. Additional noise and photoelectron-counting fluctuations would particularly affect weak neighbouring signals; thresholds remove them, while uncalibrated gain variation distorts relative amplitudes. Optical isolation would change these patterns and requires separate evaluation. Reported errors are therefore likely optimistic, and method rankings could change. Quantifying these effects requires a detector-response model with measured parameters.

Training and testing use one narrow source, with 95\% of polar angles below 10.9$^\circ$. More inclined tracks produce longer, elongated signals; lower-momentum muons alter secondary-particle activity. Wider angular or energy spectra therefore require extrapolation, which was not evaluated. The centroid also has angle-dependent bias. Experimental use requires application-matched or broader training and independent source-distribution tests.

\subsection{Implications for muon tomography}
\label{sec:tomography}

Image quality, full-system scattering-angle resolution, and material discrimination were not evaluated. Each tracking pair's direction error contributes to the measured scattering angle. For scale, the Highland formula~\cite{PDG2024} gives projected widths of 0.48$^\circ$ for 4~GeV muons through 10~cm of iron, 0.89$^\circ$ for lead, and 0.10$^\circ$ for water. At 30~cm plane separation, the centroid median space-angle error (1.90$^\circ$) exceeds these widths; transformer and CNN medians (0.35$^\circ$, 0.52$^\circ$) are comparable to iron. Space-angle errors and projected widths are different observables, so this comparison indicates scale only. Angular errors decrease approximately inversely with plane separation for fixed position errors. Learned-method tails exceed 6$^\circ$ in 1.2--2.6\% of events (figure~\ref{fig:Angle_Histogram}) and cannot be distinguished from genuine large scatters by angle alone. Establishing image-resolution or material-discrimination gains requires a full-system simulation with an object.

\section{Conclusions}
\label{sec:conclusions}

Position uncertainty in scintillator plates with pixelized readout propagates directly into the reconstructed muon trajectory and scattering angle. The energy-weighted centroid is sensitive to the computational signal grid, particularly near bin boundaries, where asymmetric collected-light patterns can shift the estimate away from the true crossing point.

All three learned methods recover sub-cell position information from the 64-cell signal maps. The transformer and CNN produce median position errors of 0.18 and 0.23~cm, compared with 0.29~cm for the MLP and 1.03~cm for the centroid. The transformer's median is approximately 2.9\% of the 6.25~cm cell width. Median space-angle errors are 0.35$^\circ$ for the transformer and 0.52$^\circ$ for the CNN, compared with 1.90$^\circ$ for the centroid and 4.21$^\circ$ for the idealized hit-cell-only reference. The transformer has the lowest median position and angular errors, while the CNN has the lowest position- and angular-RMSE point estimates. Confidence-interval overlap depends on the metric and method pair; the intervals do not include training-seed variability.

Learned reconstruction recovers sub-cell information without changing the simulated geometry. These results apply to an optically continuous plate with idealized pixelized single-count collection on all outer faces. We did not evaluate an optically isolated configuration. The gains are clearest in median and lower-percentile errors. Competing peaks consistent with secondary-particle activity produce a heavy tail, especially on the second plane, reducing the RMSE improvement.

This proof of concept uses a simplified source and effective light yield, idealized single-count optical collection on all outer faces, and no explicit photodetector or electronics model. Wider source distributions and tomographic-image quality were not evaluated. Experimental validation requires measured optical response, calibration variation, noise, thresholds, and plane-dependent effects. Tail-aware training, additional layers, or explicit treatment of secondary-electron events may reduce the tails but remain untested.

\printcredits

\section*{Declaration of competing interest}
The authors declare that they have no known competing financial interests or personal relationships that could have appeared to influence the work reported in this paper.

\section*{Data availability}
The Geant4 simulation source code, the Python analysis and reconstruction pipeline, and the derived datasets supporting this article are openly available on Zenodo at \url{https://doi.org/10.5281/zenodo.24143582}.

\section*{Acknowledgements}

This work is supported by the College of Engineering and the Grainger Institute for Engineering at the University of Wisconsin--Madison. This research was supported in part by an appointment to the Oak Ridge National Laboratory GRO Program, sponsored by the U.S. Department of Energy and administered by the Oak Ridge Institute for Science and Education.


\end{document}